\documentclass{webofc}
\usepackage[varg]{txfonts}   % Web of Conferences font
\usepackage{hyperref}
\usepackage{url}
\usepackage{float}
\hypersetup{colorlinks=true,citecolor=blue,urlcolor=blue,linkcolor=blue}

\usepackage{graphicx}

\makeatletter
\gdef\orcidlogo{%
  \includegraphics[height=1.8ex]{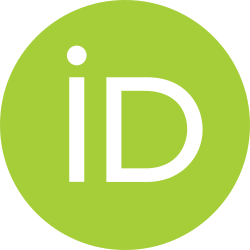}%
}
\gdef\orcid#1{\href{#1}{\orcidlogo}}
\makeatother

\newcommand{\lumiprec}{\delta\mathcal{L}/\mathcal{L}}

\begin{document}
%\linenumbers

\title{Novel Machine Learning Methods to Improve Z Pole $\quad\quad$ Integrated Luminosity at Future Colliders}

\author{\firstname{Brendon} \lastname{Madison\orcid{https://orcid.org/0000-0002-3438-4845}}\inst{1}\fnsep\thanks{\email{brendon_madison@ku.edu}}
}
% TODO: replace with final author list and affiliations.

\institute{Department of Physics and Astronomy, University of Kansas,
Lawrence, KS 66045, USA
          }

\abstract{Future $e^+e^-$ colliders at the Z pole place strong demands of $\lumiprec<10^{-4}$ on the integrated luminosity measurement. Small angle Bhabha scattering (SABS) remains the standard channel, while diphoton ($\gamma\gamma$) events provide a complementary measurement. This contribution summarizes recent work on two dominant uncertainties. First, we investigate backgrounds to the diphoton channel and find that SABS and low-invariant-mass neutral hadrons are the most significant backgrounds. A gradient boosted decision tree (BDTG) is used to classify events by particle ID. The classification results show the existing and upgraded forward tracker and luminosity calorimeter (LumiCal) designs reject neutral hadrons but only the LumiCal upgrade can reject SABS at $\lumiprec<10^{-4}$. Second, we solve the beam deflection bias problem on an event-by-event basis using two machine learning algorithms. A BDTG and the newly written Adaptive Symbolic Memetic Regression (ASMR) are trained on beam deflection data. ASMR outperforms BDTG and provides a reduced uncertainty of $5\times10^{-6}$ for beam deflection.}

\maketitle

\section{Introduction}\label{sec:intro}

Precision measurements at the Z pole remain a clear physics case for future $e^+e^-$ colliders. A Z pole run can deliver improvements in electroweak observables only if the absolute normalization of the data sample, the integrated luminosity, is controlled below the $10^{-4}$ level~\cite{MadisonLuminosityThesis,LinearColliderVision}.

We restrict this study in two ways: we consider the case of a linear collider with $(|P_{e-}|,|P_{e+}|)=(80\%,30\%)$ beam polarization, such as the International Linear Collider (ILC) or the linear collider at CERN, and we use the International Large Detector (ILD) design~\cite{ILD:MostRecent,ILD:BestCited,Behnke:ILCRef,LinearColliderVision}. We also consider an upgrade of the ILD forward region to use a highly granular calorimeter, referred to as the Granular Long Instrument for Precision (GLIP), as done in previous studies~\cite{Madison:PosRes,MadisonLuminosityThesis}.

Small-angle Bhabha scattering (SABS) remains the standard luminosity channel due to its large cross-section and well established precision theory calculation~\cite{JadachBHWIDE,JadachKKMCee}. Diphoton events, $e^+e^-\to\gamma\gamma$, are proposed to provide a complementary measurement, but with different detector and beam uncertainties and a comparably simple theory calculation~\cite{MadisonLuminosityThesis,WilsonMadison:Diphoton}. Both are restricted to the forward region, particularly the Luminosity Calorimeter (LumiCal) and forward tracker (FTD), as this leads to better uncertainties in terms of detector energy calibration and metrology~\cite{MadisonLuminosityThesis}. This paper explores two dominant uncertainties: event identification and beam electromagnetic deflection. To optimize these, this paper deploys two machine-learning algorithms: ROOT TMVA's Gradient Boosted Decision Tree (BDTG), and Adaptive Symbolic Memetic Regression (ASMR), which has been developed for this study~\cite{ROOT-TMVA}.

%Figures/ILD_forward_regionCombined_GLIPtop_ILDbottom_iFTD_singleiFTD.pdf
    \begin{figure}[t]
        \centering
        \includegraphics[width=1.0\linewidth]{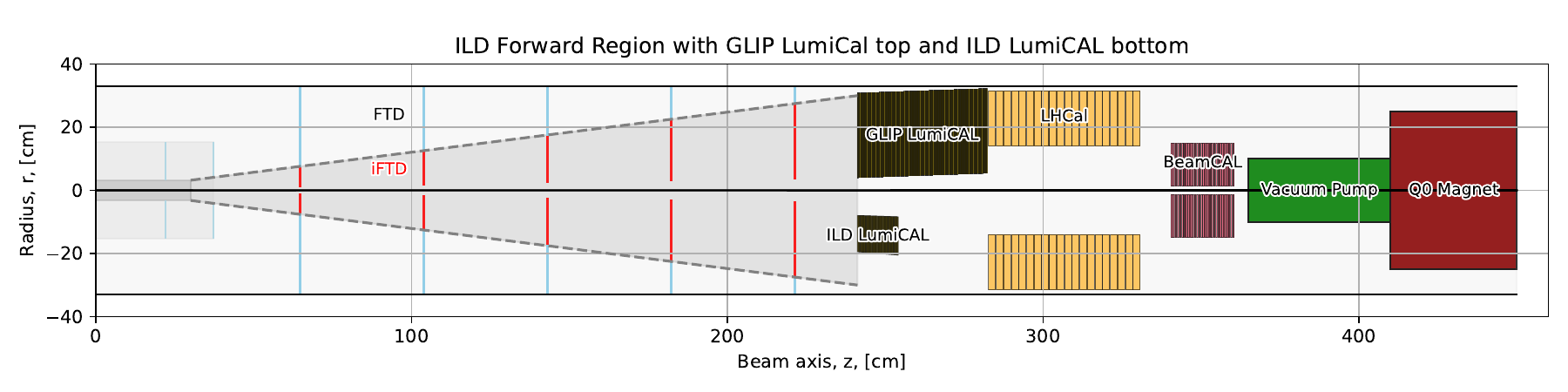}
        \caption{Forward region layout used in this study, based on the ILD forward region used at ILC~\cite{ILD:BestCited}. The existing ILD LumiCal is shown on the bottom while the GLIP LumiCal, which is further document in other studies, is shown on the top.~\cite{Madison:PosRes,MadisonLuminosityThesis}}
    \label{fig:forward_region}
   \end{figure}

For a more in-depth introduction to integrated luminosity and its measurement at future $e^+e^-$ colliders, especially at the Z pole, we recommend previous studies~\cite{MadisonLuminosityThesis,VoutsinasBeamBeam}. A diagram of the forward instrumentation can be seen in Figure~\ref{fig:forward_region}. We have extended the existing ILD FTD to the inner acceptance of the LumiCal with what we refer to as the inner FTD (iFTD) so that we can asses the performance of forward tracking over the entire acceptance of the LumiCal, instead of the existing design where the FTD acceptance would only include wider angles. The ILD LumiCal, with approximately $1X_0$ per layer, and the GLIP LumiCal, with $\frac{1}{6}X_0$ per layer, are investigated as LumiCal candidates. The upgrade to a more granular design is motivated by previous studies that show improved energy calibration, position resolution, and particle identification~\cite{MadisonLuminosityThesis,Madison:PosRes}.

\section{Backgrounds to SABS and diphotons}\label{sec:backgrounds}

For SABS, four-fermion production is the main contamination but, as reported in previous studies, this can be treated as a correction to Bhabha scattering~\cite{MadisonLuminosityThesis}. For diphotons the main contamination is from Bhabha scattering. The event identification for diphotons and Bhabha scattering was previously done, using a BDTG with both particle identification, detector and reconstruction information. This previous study found that the residual bias of Bhabha scattering contamination on the diphoton integrated luminosity precision is $35\times10^{-4}$ for the ILD LumiCal with five layer forward tracker and $0.01\times10^{-4}$ for the GLIP LumiCal with five layer forward tracker~\cite{MadisonLuminosityThesis}.

We explored numerous other possible background processes, specifically for the Z pole and diphotons. A new background process, radiative hadrons ($e^+e^-\to q\bar{q}\gamma$), where the photon is an initial state radiation (ISR) photon near beam energy and in the forward region. The recoiling hadronic system is then boosted into the opposing forward region, with an invariant mass $\lessapprox2~\mathrm{GeV}$, placing the hadronization in the realm of low energy $e^+e^-$ colliders. In this regime, radiative return to $\omega(\mathrm{782})$ or the production of $\pi^0\eta$ is common, decaying to narrow and photonic final states. These can then be confused with a single photon in the LumiCal. Using WHIZARD and the cuts from a previous integrated luminosity study, we found a cross-section of $\approx$3~pb, compared to the $\approx$100~pb cross-section of diphotons~\cite{Whizard}.

No standard simulation chain is reliable in this regime, as neither WHIZARD nor Pythia are designed to handle low invariant mass quark production and hadronization~\cite{Whizard,Pythia8}. To alleviate this, we used WHIZARD for event generation, followed by Pythia to handle parton kinematics~\cite{Whizard,Pythia8}. Existing models and data from low energy $e^+e^-$ colliders were used to hadronize partons in custom code that was added to GEANT4 to allow for the production of final states from hadronic decays~\cite{Geant4}. We find that $\approx$20\% of the $q\bar{q}\gamma$ events in the forward region can be neutral hadrons faking photons. These rates should be interpreted as order-of-magnitude estimates, as this hadronization simulation is not precise.

\begin{table}[t]
\parbox{.5\linewidth}{

\centering
\begin{small}
\caption{Particle ID confusion matrix, in percentages, for \textcolor{red}{\textbf{ILD LumiCal}} and 5 layer forward tracker.}
\begin{tabular}{lrrrr}
\hline
\textbf{True}$\rightarrow$ & $\gamma$ & hadrons & $e^-$ & $e^+$ \\
\hline
\textbf{Output}$\downarrow$& & & & \\
$\gamma$   & 96.88      & 0.48        & 1.38 & 1.26 \\
hadrons             & 0.06   & 99.78      & 0.1 & 0.06     \\
$e^-$     & 0.1   & 0.02       & 98.62 & 1.26     \\
$e^+$     & 0.14   & 0.02       & 1.16 & 98.68    \\
\hline
\label{tab:ild}
\end{tabular}
\end{small}
}
\hfill
\parbox{.5\linewidth}{

\centering
\begin{small}
\caption{Particle ID confusion matrix, in percentages, for \textcolor{red}{\textbf{GLIP LumiCal}} and 5 layer forward tracker.}
\begin{tabular}{lrrrr}
\hline
\textbf{True}$\rightarrow$& $\gamma$ & hadrons & $e^-$ & $e^+$ \\
\hline
\textbf{Output}$\downarrow$& & & & \\
$\gamma$   & 98.22      & 0.14       & 0.72 & 0.92 \\
hadrons             & 0.02   & 99.80       & 0.12 & 0.06     \\
$e^-$     & 0.16   & 0.02       & 99.24 & 0.58     \\
$e^+$     & 0.02   & 0.02       & 0.66 & 99.30     \\
\hline
\label{tab:glip}
\end{tabular}
\end{small}
}
\end{table}

A BDTG was trained on kinematic, spatial, and cluster observables from the forward calorimeter and tracker to classify photons, electrons, positrons, and neutral hadrons~\cite{ROOT-TMVA}. It does not perform event identification as done in a previous study~\cite{MadisonLuminosityThesis}. The results, seen in Tables~\ref{tab:ild} and~\ref{tab:glip}, indicate similar neutral hadron and photon separation for both LumiCal designs. The neutral hadron contamination affects integrated luminosity precision by $2\times10^{-5}$ for the ILD LumiCal and $6\times10^{-6}$ for the GLIP LumiCal. For the ILD LumiCal, the leading effect is still from misidentifying Bhabha events as diphoton events. For the GLIP LumiCal, which separates SABS and diphotons to $\approx1\times10^{-6}$, it is dominated by the neutral hadron background.

\section{Solving beam deflection bias and uncertainty}\label{sec:deflection}

Beam electromagnetic deflection occurs when the outgoing $e^+e^-$ pair from Bhabha scattering are deflected by the electromagnetic field of the bunches and their interactions. In an idealized and symmetric collider this polar angle deflection is equal and opposite, such that the electron and positron are both bent away from the beam axis and to wider angles, similar to being given an identical but small increase in transverse momentum. If left uncorrected, beam electromagnetic deflection changes the polar angle of the pairs by roughly 250 $\mu$rad at the Z pole for ILC~\cite{MadisonLuminosityThesis}. This then biases the integrated luminosity measurement by $\approx600\times10^{-4}$~\cite{MadisonLuminosityThesis}. For an introduction to this at LEP and ILC, we recommend previous studies~\cite{VoutsinasBeamBeam,MadisonLuminosityThesis}. Existing corrections to this bias are done by modeling the mean deflection using beam simulation software like GUINEA-PIG~\cite{GuineaPig}. Starting with the bias equation, Equation~3 in Voutsinas et al., and propagating uncertainty on the deflection amount, the effect on integrated luminosity precision of this approach goes as
\begin{equation}
\left(\frac{\delta\mathcal{L}}{\mathcal{L}}\right)^2 \approx \left(\frac{2}{\theta_{\min}^{-2} - \theta_{\max}^{-2}}\right)^2 
\left[\left(\frac{\sigma_{\min}}{\theta_{\min}^3\sqrt{N_{\min}}}\right)^2 + \left(\frac{\sigma_{\max}}{\theta_{\max}^3\sqrt{N_{\max}}}\right)^2\right],
\label{eq:bulk_bias}
\end{equation}
where we write the inner (min) acceptance as $\theta_{\min}$, the RMS of the distribution of the polar angle deflection amount as $\sigma_{\min}$, and the number of samples for said evaluation as $N_{\min}$, and similarly for the outer (max) acceptance we write the RMS of the distribution of the polar angle deflection amount as $\sigma_{\max}$ and the number of samples as $N_{\max}$~\cite{VoutsinasBeamBeam}.

If, instead, we have a method that can predict the amount of polar angle deflection on an event-by-event basis, then the effect on integrated luminosity comes from the residual of that prediction. If we assume this residual is dominantly Gaussian then the uncertainty is higher order, and goes as
\begin{equation}
\left(\frac{\delta\mathcal{L}}{\mathcal{L}}\right)^2
\approx
\frac{4\sigma_{\rm def}}{\sqrt{\pi}N_{\rm eve}}
\frac{
\theta_{\min}^{-3}+\theta_{\max}^{-3}
}{
(\theta_{\min}^{-2}-\theta_{\max}^{-2})
},
\label{eq:ind_bias}
\end{equation}
where $N_{\rm eve}$ is our total number of events we train and test on, we multiply by two to reflect cross-validation, and we have $\sigma_{\mathrm{def}}$, which comes from the RMS of the residual of the reconstruction of the polar angle deflection with the true value of the polar angle deflection.

\section{Machine learning regression of beam deflection}\label{sec:mlregression}

For brevity, we restrict ourselves to solving for the electron event-by-event deflection. Two machine learning regression algorithms were used: a BDTG from ROOT's TMVA library and an algorithm developed for this study, ASMR. The latter, being symbolic memetic regression, is fundamentally different from a BDTG, as it can learn symbolic, possibly analytic, solutions to the problem. Since it is memetic, it uses a suite of local optimization algorithms, including Minuit, in addition to its global evolutionary scheme~\cite{Minuit}. ASMR is also adaptive in that it can change the partitioning of the data it is training on, using equipartioning of the data, and it can schedule various reward schemes and temperatures, scaling the randomness of its candidate models. Both models were restricted to using the GLIP LumiCal, which has an energy resolution of $\approx\frac{5.5\%}{\sqrt{E}}$ and a polar angle resolution of $\approx10$ $\mu$rad~\cite{MadisonLuminosityThesis}. The test and training datasets were generated using BHWIDE and GUINEAPIG, to simulate SABS event generation and beam electromagnetic deflection respectively, and then a detector fast simulation was used to Gaussian smear the results according to the forward region given previous GEANT4 simulations~\cite{JadachBHWIDE,GuineaPig,Geant4,MadisonLuminosityThesis}. We then used the reconstructed energies and polar angles of the SABS $e^+e^-$ pair, represented as $(E_+,E_-,\theta_+,\theta_-)$, as the input independent variables, and the electron polar angle at the event generator level, from BHWIDE and before the deflection simulated in GUINEAPIG, as the dependent variable~\cite{JadachBHWIDE,GuineaPig}.

\begin{figure}[t]
        \centering
        \includegraphics[width=0.5\linewidth]{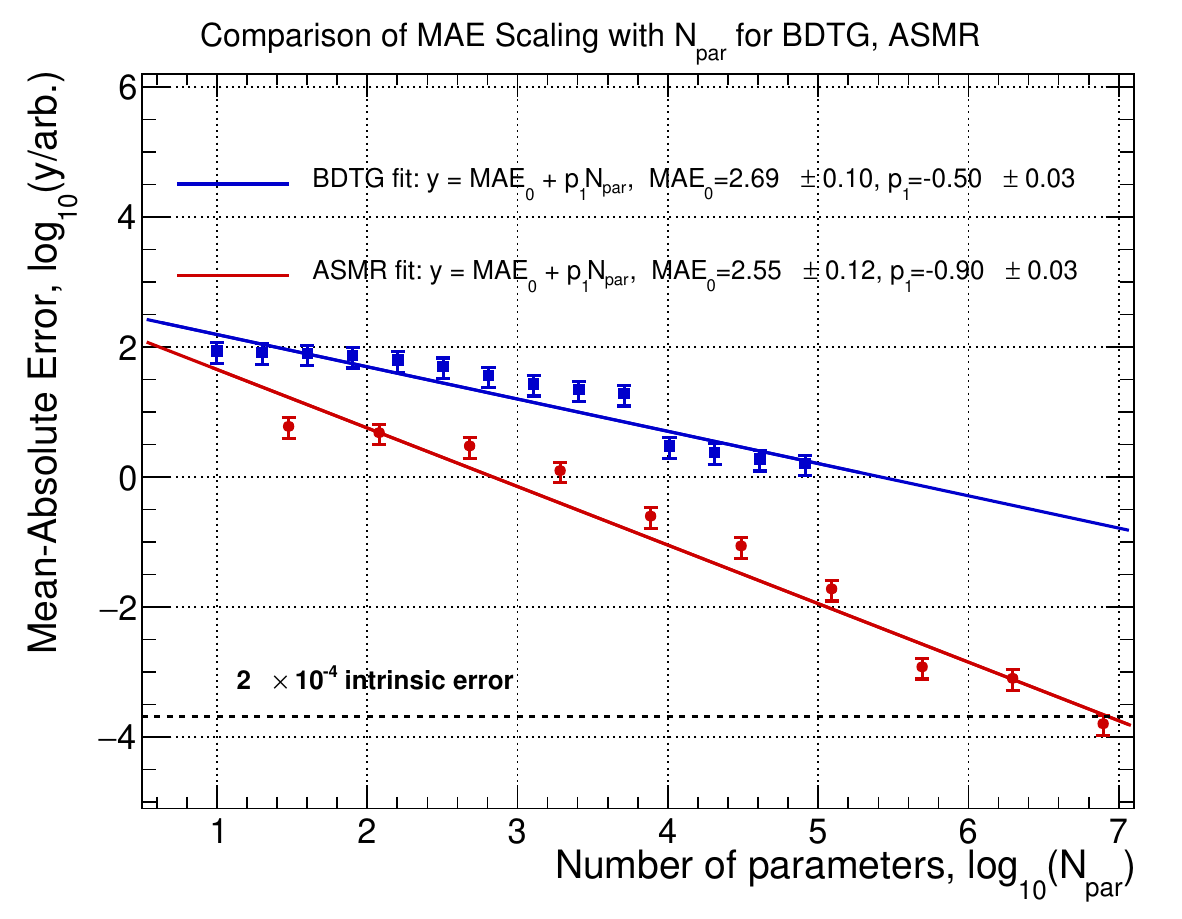}
        \caption{MAE scaling of BDTG and ASMR with parameter count for the benchmark test that used the W-M function, with arbitrary units and $2\times10^{-4}$ intrinsic error added using Gaussian smearing.}
    \label{fig:scaling}
   \end{figure}

During this study, the machine learning algorithms were first benchmarked using the W-M function, which is a fractal with tunable dimensionality~\cite{WMFunc}. The performance of each algorithm is evaluated using the mean-absolute error (MAE). The performance on the benchmark can be seen in Figure~\ref{fig:scaling}. During training, ASMR managed to learn to construct Fourier series as solutions to the W-M function, which is a known approximation.

\begin{figure}[t]
        \centering
        \includegraphics[width=0.5\linewidth]{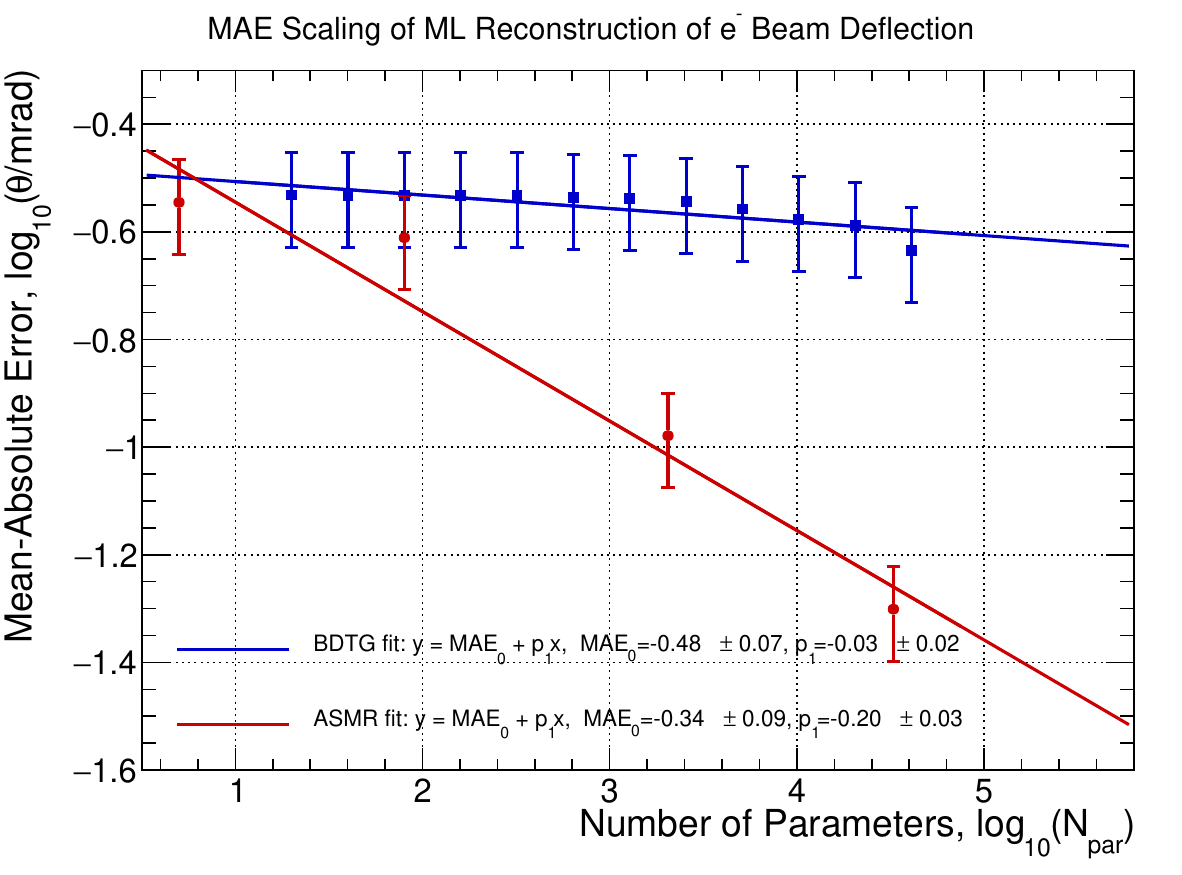}
        \caption{Regression performance for the event-by-event beam deflection reconstruction of the outgoing electron using the BDTG and ASMR algorithms.}
    \label{fig:deflection_performance}
   \end{figure}

After benchmarking, BDTG and ASMR were unblinded and run on the electron deflection data, with performance shown in Figure~\ref{fig:deflection_performance}. In both cases, ASMR out-performs BDTG in terms of raw MAE, as well as scaling with the number of parameters used in the solution. During training on the electron deflection dataset, ASMR derived and learned that $\frac{\theta_--\theta_{+}}{2}$ is a good leading order predictor of the amount of deflection. This agrees with the back-to-back case where the deflection is equal and opposite. When this variable and other common kinematic variables were provided to the BDTG, it found $\frac{\theta_--\theta_{+}}{2}$ to be the highest in terms of variable importance.

For the 200k events trained on here, with half being used for cross-validation, we observed a best case performance of $\sigma_{\mathrm{def}}\approx$ 50 $\mu$rad from the ASMR algorithm. This results in an effect of $5\times10^{-6}$ on integrated luminosity precision, which we will treat as the same for both LumiCal designs. The mean based method for the same 200k event data sample finds $8\times10^{-5}$. Showing that the event-by-event method is more precise.

\section{Updated uncertainty picture and outlook}\label{sec:outlook}

Recent updates on beam polarization error propagation for diphotons, including higher order corrections, have reduced the effect of the beam polarization uncertainty~\cite{Madison:Polarization}. This reduction, combined with using both beam polarimeter and event fitting measurements of beam polarization to reach $10^{-3}$ on beam polarization precision, further reduces the effect on integrated luminosity to $4\times10^{-4}$.

The results from particle and event identification clarify that SABS contamination of diphoton events is still the largest hurdle for the existing ILD LumiCal design. Only with an upgrade to the GLIP LumiCal does the contamination of SABS become less than that of neutral hadrons from $q\bar{q}\gamma$, with both at the $10^{-6}$ level. This makes the case for an upgraded LumiCal clear, as any experiment that wishes to use diphotons must use a design with upgraded granularity that can reject SABS well enough and then, at even higher precision, reject neutral hadrons. Though more precise modeling, and potentially a dedicated simulator, of the neutral hadron channel is needed.

With the machine learning algorithms offering event-by-event solutions to the beam deflection amount, the associated effect on integrated luminosity has significantly decreased. We have also shown that the event-by-event solution is likely more precise than the average driven method. This comes with the assumption that one can model the underlying physics to a similar level as $5\times10^{-6}$ with the methods used.

The use of both diphotons and SABS is even more important when one uses modeling and machine learning approaches to minimize beam deflection bias because, if the models and solutions are inaccurate, there would be disagreement between the two measurements. This allows a natural way to verify the accuracy and correct the models and solutions, while also providing for a combined measurement of integrated luminosity that out-performs the two individual measurements, as seen in Table~\ref{tab-LumiTableZ}.

Beyond the recent work done here and elsewhere, there is still a significant need for improvements in the angular resolution and precision of beam polarization measurement. These provide the dominant out-standing sources of uncertainty not addressed in this work. Until these improvements are provided, it is unknown whether a future $e^+e^-$ collider can surpass $1\times10^{-4}$ at the Z pole. Even at an unpolarized collider, the angular resolution will dominate above this level. Funding for future studies and further LumiCal development is needed to allow for the incredible Z pole electroweak precision measurements that future $e^+e^-$ colliders plan to conduct.

\begin{table}[t]
\centering
\caption{Estimated integrated luminosity uncertainties using small-angle Bhabhas and diphotons at the Z pole. Measured in the FTD and the listed luminosity calorimeter. For more details see other studies~\cite{MadisonLuminosityThesis}. All numerical values are in parts per ten thousand ($10^{-4}$) and assumed uncorrelated.\begin{tiny}\\ *Dominated by neutral hadron contamination, which needs further study to improve accuracy of estimate\\
$^\dagger$Using acceptance and angular resolution from ILD LumiCal (22 $\mu$rad) and GLIP LumiCal (3 $\mu$rad) studies~\cite{Madison:PosRes,ILD:PosRes}
\end{tiny}}
\begin{tabular}{|l|c|c|c|c|}
\hline
\textbf{Uncertainties} & \textbf{LumiCal SABS} & \textbf{LumiCal $\gamma\gamma$} & \textbf{GLIP SABS} & \textbf{GLIP $\gamma\gamma$} \\
\hline
Beam Deflection Effects & 0.05 & -- & 0.05 & --  \\
Polarization of beams & 0.01 & 4.0 & 0.01 & 4.0  \\
Statistics (for 0.1 $\mathrm{ab}^{-1}$) & 0.10 & 5.0 & 0.10 & 3.5 \\
Angular resolution$^\dagger$ & 17 & 12 & 2.3 & 2.1 \\
Energy resolution bias & 1.6 & 1.6 & 0.10 & 0.10 \\
Depth position bias & 2.6 & 2.6 & 0.30 & 0.30 \\
PID and Contamination & 0.01 & 35 & $\approx0$ & 0.06* \\
\hline
\textbf{Sum : } & \textbf{17} & \textbf{38} & \textbf{2.3} & \textbf{5.7} \\
\textbf{Combined: } & & \textbf{16} & & \textbf{2.2} \\
\hline
\end{tabular}
\label{tab-LumiTableZ}
\end{table}

\global\long\def\bibname{References}%

\bibliographystyle{JHEP}
\bibliography{lcws25}

\end{document}